**Putting the astronomy back into Greek calendrics: the *parapegma* of Euktemon**

Robert Hannah (University of Waikato)

It is a pleasure to be able to offer a paper to our honorand. Many years ago James Evans established himself as a great teacher of the history of ancient Greek astronomy to many beyond the confines of his own lecture room through his book, *The History and Practice of Ancient Astronomy*. While in more recent years he has provided us with sophisticated papers on the more technical aspects of astronomy, especially as they pertain to the Antikythera Mechanism, it to that earlier monograph, and its impact on myself and my own students, that I wish to pay homage in this small offering on ancient 'observational' astronomy.

*Parapegmata*

Calendars across all cultures in the world have traditionally relied on the observed and measured motions of the celestial bodies: the sun, the moon, and the stars. From an early stage in historical Greek society the motions of the stars in particular played a significant part in time-measurement and the development of calendars. Alongside the phases of the moon (the fundamental basis of all Greek civil calendars) and the apparent movement of the sun (which provided seasonal markers), the Greeks also used the appearance or disappearance of stars at dawn or dusk to establish schedules for timekeeping.[1]

While we are aware of these uses of astronomy from the poems of Homer and Hesiod, it is from the fifth century BC onwards that Greek astronomers formalised such star data into *parapegmata*. These survive from later periods as stone tablets inscribed with day-by-day entries for the appearance or disappearance of stars.[2] We may regard these as 'star schedules', or more loosely as 'star calendars'. These were set up in public spaces in the cities, and so had a civic significance beyond the narrowly astronomical, much as clocks did in Medieval and Early Modern Europe. Many of the leading astronomers of antiquity played a role in the development of the *parapegmata* – the list includes Meton (inventor of the 19-year 'Metonic Cycle', which still governs the placement of Christian Easter in the western calendar, and the Jewish calendar as a whole), Euktemon, Eudoxos, Kallippos, and Ptolemy.[3] *Parapegmata* continued in use until the Medieval period.[4] In literary form, they were combined into compilations and published either in their own right by astronomers (e.g. in Geminos, *Eisagoge*; Ptolemy, *Phaseis*), or subsumed into agricultural 'handbooks' by literary authors, especially by the Romans (e.g. Columella, *On Agriculture*; Varro, *On Farming*). Star lore of this kind pervades every aspect of Greek and Roman literature: it can be found in all the major authors, from Aeschylus to Euripides and Aristophanes, through Aratos to Plautus, Vergil and Ovid and beyond. Julius Caesar was credited with a 'star calendar', which survives in later quotations (Pliny, *Natural History* 1 and 18). In all, about 60 *parapegmata* survive in epigraphical and literary form.[5]

Despite the pervasiveness of such astronomical data in the broader culture of Classical antiquity, this 'star calendar' information has been regarded as inaccurate in itself or in the uses to which it was applied. This issue goes back to antiquity – Pliny the Elder (*Natural History* 18. 210–213) complained about the different dates given in his sources for the same star phenomena when observed in the same country. The complaint is picked up in modern scholarship.[6] This has been especially the case with regard to the dates for the phenomena given by the Roman poet Ovid in his calendar-poem, the *Fasti*.[7] But recent studies suggest that the astronomical basis of the star data relayed by Ovid has been misunderstood, and that modern parameters for accuracy have been imposed anachronistically on the ancient data.[8]

The *parapegma* was first dealt with in any significant manner in a series of publications by Rehm in the first half of the 20[th] century.[9] Several papers by van der Waerden treated the topic in the second half of the century,[10] but because he based his work on Rehm's, it is still the latter's views which lie behind his scholarship. Unfortunately, as others have noted, Rehm mixed fact and hypothesis indiscriminately in his conclusions,[11] and a rigorous critique of his work shows that there is a need to re-lay the astronomical foundations for work on the subject. Furthermore, Rehm and van der Waerden were primarily interested in simply reconstructing the *parapegmata* as lists of observations, rather than situating them into a wider context. The most recent, and now fundamental, book specifically on *parapegmata* still avoids any treatment of their astronomy in favour of dealing with them as 'omen literature'.[12] This is a valid enterprise, and one to which, in some respects, this paper may contribute, although that was not its aim. In this paper I wish to rehabilitate the 'star calendar' to its proper place in the history of science, as a time-marking instrument which once pervaded many aspects of Greek and Roman culture, and which was accurate within the parameters of its time.

A *parapegma* afforded the facility to measure time through the year via the stars. *Parapegmata* were set up in public areas of Greek cities. But Greek city-states regularly used the moon as the basis for their calendars, not the sun or the stars. Because the moon's cycle is incommensurate with the sun's, and therefore with the seasons, these city calendars did not sit well with the seasons, and needed considerable adjustment to bring, for instance, agricultural festivals back into synchronisation with the appropriate season.[13] (We still do this nowadays with Jewish Passover or Christian Easter and the northern hemisphere season of spring.) But the cycle of the stars corresponds very closely with the sun's, and so following the stars as timekeepers effectively means, over a person's lifetime, following the sun. Therefore 'star calendars' can reasonably be seen as analogues and precursors to a true solar calendar. And indeed, when Greece did finally adopt a solar calendar, in the form of the Julian calendar in the time of the emperor Augustus, it had no difficulty in doing so, because it was already familiar with a very similar cycle through its use of 'star calendars'.[14]

But I have also argued elsewhere that civic authorities could have used the *parapegmata* as regulators for their ordinary lunar calendars, along with lunisolar cycles like the 'Metonic' cycle, which allowed civic authorities to synchronise the moon's and sun's cycles over repeated periods of 19 years.[15] In this context, I would suggest, the placement of *parapegmata* in public contexts makes sense. But to substantiate this claim, the accuracy of *parapegmata* as astronomical instruments must be established.

Using Pliny, *Natural History* 18. 218, Matthew Fox has recently posited that ancient observations of the stars took the sun to be at a certain distance below the horizon, apparently regardless of the brightness (magnitude) of the star being observed on the horizon.[16] Pliny says the sun must be 'at least three-quarters of an hour' below the horizon. Le Bonniec and Le Boeuffle suggest that this corresponds to about 12° below the horizon.[17] Hitherto, it had been assumed that ancient observations, like all modern ones, took the magnitude of each star into account, and therefore required the sun to be at differing distances below the horizon.[18] The star Sirius, for example, has a magnitude of -1.44 and is the brightest star in the sky. Modern calculations of the sun's minimum distance below the horizon assume only 7° for observations of Sirius when it is rising just after sunset, or setting just before sunrise. In Athens in 430 BC, the calculated date for the morning setting of Sirius, with the dawning sun only 7° below the horizon, is 24 November in our terms. A depression of 12° for the sun below the horizon causes several days' difference in the calculated observation time, shifting it forward to 29 November. In Euktemon's *parapegma*, this same event is dated slightly later again, to the equivalent of 1 December (following Rehm's reconstruction), arguably supporting the supposition that the sun is further below the horizon than just the 7° that modern calculations assume on the basis of the star's brightness.

The assumption that magnitude mattered (as it does to us today) can therefore significantly affect the calculated dates of the original observations by several days, hence the modern assumption of inaccuracy in the ancient *parapegmata* or their secondary users, like the poet Ovid. The removal of magnitude from the observation equation represents a major shift in our thinking, the broader significance of which has yet to be understood.

In addition, Pliny's use of a time period (three-quarters of an hour) outside daylight hours must signify the adoption of a water-clock, and therefore implies the prioritisation of an external timing device over optical observation of the star, i.e. the astronomer was not waiting night after night to see the first or last appearance of a star on the dim horizon, but was using a timer to tell him when to look – I am tempted to add, if he looked at all, as the late Douglas Kidd once said to me. So there was a mechanical artifice governing the so-called 'first' and 'last' observations. Modern mathematical calculations ignore this external imposition, and in addition assume a clear and flat horizon. From the time of Eudoxos in the fourth century BC, we might suppose use of some form of celestial globe.[19] Even some small time before then, however, Plato's description of the cosmos in his *Timaeus* suggests that something like the more skeletal armillary sphere was familiar to him.[20] We have no physical evidence of such instruments from this period, so this can only be an educated piece of speculation for the moment.

*Euktemon's parapegma*

In what follows I deal only with a small section of the data set preserved from Euktemon's *parapegma*, namely the first part of the summer section from the morning rising of the Pleiades to the morning rise of Sirius. It is enough to give a flavour of the type of work which might be pursued further; the relative strength of the argument regarding a time-limit rather than pure ocular observation; and certainly one of the apparent weaknesses of whatever instrumentation was used to time the 'observations'. The Greek text followed here is Aujac's; the translation is my own.[21] The astronomical data (RA, Dec, magnitude) are derived from the computer planetarium programme, Voyager 4.5.[22] Dates for the various stellar phases – 'visible' (unless otherwise stated) morning rising, morning setting, evening rising, evening setting (taking Evans' point that the modern terminology of heliacal rising and setting, achronychal rising and cosmical setting can be unhelpful and is largely anachronistic)[23] – are derived from my own calculations based on the trigonometrical formulae which must underlie the examples in Neugebauer 1925; I have added here in Appendix 1 my sample derivation of Neugebauer's method, since the trigonometry behind it seems largely unknown now to archaeoastronomers. I have entered the data derived from these calculations to the Voyager planetarium programme to gain a visual impression. These readings have been set there against an horizon of Athens as seen from the ancient Pnyx, where Meton, Euktemon's colleague, was said to have set up a *heliotropion*, and which has largely not changed since antiquity. As it turns out, however, the readings suggest strongly that this was not an observation point. A more mathematically precise methodology for star observations has been presented by Schaefer, and used by, e.g., Salt and Boutsikas.[24] This, however, is over-precise and hence illusory or misleading, since it overrides the very human, subjective process of observation. Anyone who has tried to see a 'first' or 'last' phenomenon (as Schaefer himself said he had) will know that it applies only to himself or herself; one's very neighbour may not see it.

[Τὸν δὲ Ταῦρον διαπορεύεται ὁ ἥλιος ἐν ἡμέραις λβ.

The sun passes through Taurus in 32 days.]

> This could not be in Euktemon's *parapegma*, but the zodiac is used in Geminos as the organising principle; the usage stems probably from Kallippos's *parapegma*.[25]

Ἐν δὲ τῇ ιγ^η Εὐκτήμονι Πλειὰς ἐπιτέλλει· θέρους ἀρχή· καὶ ἐπισημαίνει.

Day 13 (May 7), according to Euktemon Pleiades rise (Pleiades morning rising); beginning of summer; and there is sign of weather.[26]

| STAR | RA | Dec | Mag | Geminos | Hannah |
|------|-----|-----|-----|---------|--------|
| η Tau | 01h31m30.7s =22˚.65 | +13˚46'16.5" =+13˚.83 | 2.86 | 13 Taurus =7 May | 23 May |

On 7 May when η Tau is rising, the sun is 9˚29' below the horizon. This is technically not 'visible' – with a flat horizon first visibility would occur ca. 23 May, with the sun 16˚ below the horizon. But a date of 7 May puts the sun 52 (equinoctial) minutes short of rising, which practically suits Pliny's criterion.

If the viewpoint was the Pnyx, then η Tau rises between the Acropolis and Mt Lykabettos, at an altitude of about 6˚ and an azimuth of 65˚07', and is first visible about 16 June, with the sun 16˚ below a hypothetical flat horizon, but about 22˚ below the actual hilly horizon – as per Boutsikas and Salt at Delphi, I am retaining the notional angle of depression for a flat horizon observation (although I think this is unlikely to be the real angle).[27] This later date would count against viewing from the Pnyx. See Figure 1.

Ἐν δὲ τῇ λα^η Εὐκτήμονι Ἀετὸς ἑσπέριος ἐπιτέλλει.

Day 31 (May 25), according to Euktemon Eagle rises in the evening (Aquila evening rising).[28]

| STAR | RA | Dec | Mag | Geminos | Hannah |
|---|---|---|---|---|---|
| α Aql | 17h53m27.1s =268˚.36 | +05˚50'30.8" =+05˚.842 | 0.93 | 25 May | 25 May |

On 25 May when α Aql is rising, the sun is 7˚10' below the horizon. This is 'visible' and accurate –with a flat horizon last visibility would occur on 25 May, with the sun 7˚ below the horizon. The date of 25 May puts the sun 41 (equinoctial) minutes after setting, which suits Pliny's criterion.

Ἐν δὲ τῇ λα[η] Εὐκτήμονι Ἀρκτοῦρος ἑῷος δύνει· ἐπισημαίνει. … Εὐκτήμονι Ὑάδες ἑῷαι ἐπιτέλλουσιν· ἐπισημαίνει.

Day 32 (May 26), according to Euktemon Arktouros sets at dawn (Arcturus morning setting); there is sign of weather. ... Hyades rise at dawn (Hyades morning rising); there is sign of weather.

| STAR | RA | Dec | Mag | Geminos | Hannah |
|---|---|---|---|---|---|
| α Boo | 12h20m6.13s =185˚.03 | +31˚51'6.2" =+31˚.85 | 0.15 | 26 May | 5 June |
| α Tau | 02h22m26.27s =35˚.61 | +08˚15'47.2" =+08˚.3 | 0.99 | 26 May | 4 June |

On 26 May when α Boo is setting, the sun is 2˚08' below the horizon. This is technically not 'visible' – with a flat horizon first visibility would occur ca. 4 June, with the sun 7˚ below the horizon. A date of 26 May puts the sun only 12 (equinoctial) minutes short of rising, which does not suit Pliny's criterion. Technically, this counts as a true, and therefore invisible, morning setting. As we shall see elsewhere, however, morning settings are characteristically not this *parapegma*'s forte, so I wonder what that tells us of the instrumentation used to establish the date.

On 26 May when α Tau is rising, the sun is 6˚38' below the horizon. This is technically not 'visible' – with a flat horizon first visibility would occur ca. 4 June, with the sun 11˚ below the horizon. But a date of 26 May puts the sun 38 (equinoctial) minutes short of rising, which practically suits Pliny's criterion.

[Τοὺς δὲ Διδύμους ὁ ἥλιος διαπορεύεται ἐν ἡμέραις λβ.
The sun passes through Gemini in 32 days.]

Ἐν δὲ τῇ κδ^η Εὐκτήμονι Ὠρίονος ὦμος ἐπιτέλλει.
Day 24 (June 19), according to Euktemon shoulder of Orion rises (Orion morning rising).

| STAR | RA | Dec | Mag | Geminos | Hannah |
|------|-----|-----|-----|---------|--------|
| γ Ori | 03h17m59.27s | +0˚43'32.4" | 1.65 | 19 June | 30 June |

|  | =49˚.5 | =+0˚.73 |  |  |  |

Orion's shoulders are γ Ori (upper, so seen first) and α Ori (lower, so second).[29] According to calculation, the visible morning rising of both is on the same date (30 June).

On 19 June when γ Ori is rising, the sun is 7˚00' below the horizon. This is technically not 'visible' – with a flat horizon first visibility would occur ca. 30 June, with the sun 14˚ below the horizon. But a date of 19 June puts the sun 42 (equinoctial) minutes short of rising, which practically suits Pliny's criterion.

[Καρκίνον διαπορεύεται ὁ ἥλιος ἐν ἡμέραις λα.
The sun passes through Cancer in 31 days]

[<Ἐν μὲν οὖν ˰τῇ> α^η ἡμέρᾳ Καλλίππῳ Καρκίνος ἄρχεται ἀνατέλλειν· τροπαὶ θεριναί· καὶ ἐπισημαίναι.
Day 1 (June 28). according to Kallippos, Cancer begins to rise; summer solstice; and there is sign of weather.]

<Ἐν δὲ τῇ> ιγ^η ἡμέρᾳ Εὐκτήμονι Ὠρίων ὅλος ἐπιτέλλει.
Day 13 (July 10), according to Euktemon all of Orion rises (Orion morning rising).

| STAR | RA | Dec | Mag | Geminos | Hannah |
|---|---|---|---|---|---|
| κ Ori | 03h54m17.39s<br>=58˚.572 | -13˚40'47.6"<br>=-13˚.667 | 2.05 | 13 Cancer<br>=10 July | 20 July |

I assume that 'all of Orion' means not just β Ori but also κ Ori, as these are the 'two feet' of Orion in Aratos (*Phaen*. 338).

On 10 July when κ Ori is rising, the sun is 7˚55' below the horizon. This is technically not 'visible' – with a flat horizon first visibility would occur ca. 20 July, with the sun 14˚ below the horizon. But a date of 10 July puts the sun 48 (equinoctial) minutes short of rising, which suits Pliny's criterion.

<Ἐν δὲ τῇ> κζ^η Εὐκτήμονι Κύων ἐπιτέλλει.
Day 27 (July 24), according to Euktemon Dog rises (Sirius morning rising).

| STAR | RA | Dec | Mag | Geminos | Hannah |
|---|---|---|---|---|---|
| α CanMaj | 04h57m30.12s<br>=74˚.376 | -17˚14'26.6"<br>=-17˚.241 | -1.44 | 27 Cancer<br>= 24 July | 31 July |

I assume 'Dog' means α CanMaj, not β CanMaj.

On 24 July when α CanMaj is rising, the sun is 6˚26' below the horizon. This is technically not 'visible' – with a flat horizon first visibility would occur ca. 31 July, with the sun 11˚ below the horizon. But a date of 24 July puts the sun 38 (equinoctial) minutes short of rising, which practically suits Pliny's criterion.

<Ἐν δὲ τῇ> κη' Εὐκτήμονι Ἀετὸς ἑῷος δύνει· χειμὼν κατὰ θάλασσαν ἐπιγίνεται.
Day 28 (July 25), according to Euktemon Eagle sets at dawn (Aquila morning setting); storm at sea comes on.

| STAR | RA | Dec | Mag | Geminos | Hannah |
|------|------|------|------|---------|--------|
| α Aql | 17h53m27.76s =268˚.37 | +05˚50'31.2" =+05˚.842 | 0.93 | 25 July | 30 July |

I assume 'Eagle' means α Aql (the last part of Aquila to set), not λ Aql (the first part), whose setting date is much earlier.

On 25 July when α Aql is setting, the sun is 3˚14' below the horizon. This is technically not 'visible' – with a flat horizon first visibility would occur ca. 30 July, with the sun 7˚ below the horizon. A date of 25 July puts the sun 19 (equinoctial) minutes short of rising, which does not suit Pliny's criterion. But on 30 July the sun would be 42 (equinoctial) minutes short of rising, which would suit Pliny's criterion. As we have seen already, however, morning settings are a weak point in this *parapegma*.

[Τὸν δὲ Λέοντα διαπορεύεται ὁ ἥλιος ἐν ἡμέραις λα.
The sun passes through Leo in 31 days.]

Ἐν δὲ τῇ αῃ ἡμέρᾳ Εὐκτήμονι Κύων μὲν ἐκφανής, πνῖγος δὲ ἐπιγίνεται· ἐπισημαίνει. Day 1 (July 29), according to Euktemon Dog is visible (Sirius morning rising), and stifling heat comes on; there are signs of weather.

| STAR | RA | Dec | Mag | Geminos | Hannah |
|------|-----|------|-----|---------|--------|
| α CanMaj | 04h57m30.15s =74˚.376 | -17˚14'26.4" =-17˚.241 | -1.44 | 1 Leo = 29 July | 31 July |

I assume 'Dog' means α CanMaj.

On 29 July when α CanMaj is rising, the sun is 10˚21' below the horizon. This is technically just 'visible' – with a flat horizon first visibility would occur ca. 31 July, with the sun 11˚ below the horizon. This date of 29 July puts the sun 61 (equinoctial) minutes short of rising, which overshoots Pliny's criterion. Evans and Berggren suggest that the two 'observations' for the Dog signify a first fleeting visibility (27 Cancer / 24 July) and then an easy visibility (this present date).

Throughout this brief commentary on the star phases I have referred to Pliny's criterion of three-quarters of an hour. This risks being anachronistic, unless Pliny simply used a criterion handed down from older times and translated it into contemporary terms. If we were to consider the time devices available to Euktemon, however, it is possible that a bucket, or several buckets, could have been used to measure the time for each 'observation'. Different measures were used for different types of court cases in ancient Athens. Ten *choes*, for example, measured out cases involving more than 5,000 drachmas and seven *choes* those under that amount. A two-*choes* clay bucket, excavated in the Athenian Agora, was found to empty out in six minutes.[30] Seven *choes* – a culturally recognised amount – could empty out in 21 minutes, and two such periods would make 42 minutes, a span of time close both to Pliny's criterion and to the time-periods noted in the star 'observations' above between star-rise/star-set and sunrise/sunset. But why this particular period of time would be aimed at remains unknown.

# APPENDIX
P. V. Neugebauer's method for calculating the visibility of stellar phenomena
(Neugebauer 1925)

| **NEUGEBAUER'S TEXT** | **MY COMMENTS** |
|---|---|
| page XXXVIII: Beispiel | |
| 24a)   501 v.Chr. = -500 | The given year (501 BC) is transformed into the astronomical year. |
| 24b)   Sirius -500: | The **equatorial** coordinates for Sirius in 501BC. |
|         a = 73°.7 | a = Right Ascension, the distance in longitude from the vernal equinox (a) to the star |
|         d = -16°.4 | d = Declination, the distance in latitude above or below the equator. |
| 24c)   l = 15° östlich Greenwich,<br>         f = 34° | l = geographical longitude<br>f = geographical latitude of the observer (15°E, +34°). |
| 24d)   Sirius ist ein Stern 1. Grösse.<br>         Tafel 28.<br>         b = 11 für heliakischen<br>         Aufgang und Untergang<br>         b = 7 für scheinbaren<br>         akronychischen Aufgang<br>         und scheinbaren kosmischen<br>         Untergang. | Sirius is a 1$^{st}$ magnitude star<br>b = -11° for arc of visibility below horizon at heliacal rising and setting<br>b = -7° for apparent acronychal rising and apparent cosmical setting.<br>= sun's minimum negative altitude (in **horizon** coordinates) below the horizon for the star still to be visible just above the horizon. |

24e) Tafel 1.
Vertikal Argument
d = -16˚.4
Horizontales Argument
f = 34˚.0
t = 5$^h$.29
oder in Grad = 79˚.4

Formula:

t = cos-1{-tan f tan d}
= the semi-diurnal arc, the time or distance between the rising or setting of a star on the horizon and its transit over the observer's meridian.
= the hour angle (H) of a star at rising or setting.
Hour angle (H) = how far the star has travelled since it crossed the meridian, so here t = time or distance, expressed in **equatorial** terms, between the meridian and the rising/setting point of the star on the horizon.
The star's zenith distance = 90˚ at its rising/setting point, but this is in altazimuth terms, not equatorial.

Following the calculations for heliacal rising and apparent acronychal rising:

a = 73°.7
t = 79°.4
y = a - t = 354°.3

y = the time elapsed/distance covered since the Vernal Equinox (VE) last crossed the meridian, and therefore also = where the meridian is. Whenever Sirius rises, at any time of the day, VE lies about 24 minutes (0.4h) east of the meridian (i.e. a will cross the meridian about 24 minutes after Sirius rises). VE's precise position at Sirius's rising corresponds to the difference between the semi-diurnal arc of Sirius (t = 79°.4) and the distance between Sirius and VE (= R.A. of Sirius, a = 73°.7) = -5°.7 (= the approx. 24 minutes between VE and the meridian), to which 360° must be added to make it positive, i.e.
y = 360° - (t - a) = 360° - t + a = 360° + a - t = 360° + 73°.7 - 79°.4 = 360° - 5°.7 = 354°.3 (= 23h 37m 12s R.A.), which is effectively Neugebauer's formula.
[In the case of calculating heliacal / cosmical settings, y = t + a = the sum of the semi-diurnal arc (t) – which takes us from the meridian westwards to the point of setting for Sirius – plus the R.A. of Sirius – which takes us beyond Sirius round to where a now lies; i.e.
y = 79°.4 + 73°.7 = 153°1 = 10h 12m 24s R.A.]
This means that the meridian lies at 354°.3 R.A. in the case of Sirius's rising, and at 153°.1 in the case of its setting.
In its turn, the R.A. of the meridian is also necessarily the R.A. of the zenith, since it lies on the meridian.

24f)   Argument y = 354˚.3:
       p = -0˚.6
       r = +2˚.6
       s = +23˚.6
       S = f + r = +36˚.6

24g)   Tafel 25.
       Vertik. Arg. s = +23˚.6
       Horiz. Arg. S = +36˚.6
       P = +16˚.6
       Tafel 26.
       Dieselben Argumente:
B = +33˚.2
y = 354˚.3
p = -0˚.6
P = +16˚.6
L = 10˚.3

24f) - 24g):
y = R.A. of the meridian, and
f = observer's geographical latitude.

Formulae:
L = tan$^{-1}$ {(sin y cos e + tan f sin e) / cos y}
B = sin$^{-1}$ {sin f cos e - cos f sin e sin y}
where e = the obliquity of the ecliptic.

y and f are converted from equatorial to **ecliptic** form (L, B).

f by extension = Declination of the zenith point.
Converting f into ecliptic form (B) not only gives the latitude of the zenith point from the ecliptic, but more importantly the angle below the ecliptic, between it and the horizon (= 90˚ - B, since zenith-to-horizon = 90˚). This will be used to calculate L$_1$: see 24h) below.

y is converted to ecliptic form to provide the sun's position on the observer's meridian (the sun's ecliptic longitude, L), when Sirius is rising or setting.
We need to maintain this angular relationship between Sirius and the sun. Putting the sun on the meridian as Sirius rises or sets gives a time (noon) and date when the sun and Sirius are in this angular relationship. But it is not the date of Sirius's HR, AR, HS or CS, because if we then move the sun of this date to either the eastern or the western horizon, Sirius will also have moved commensurately forwards or backwards from its position on the horizon, to maintain the angle between it and the sun.

We therefore still need to find when the sun is on or just below the horizon as Sirius rises or sets.

24h) Heliakischer Aufgang  
    b = 11˚  
    Tafel 27.  
    Spalte b = 11˚  
    Arg. B = 33˚.2  
    $L_1$ = 13˚.2

Heliacal Rising  
The formula at Tafel 27 is:  
$\sin L_1 = (\sin b / \cos B)$,  
so $L_1 = \sin^{-1}(\sin b / \cos B)$  
$L_1$ = the angular distance on the ecliptic from the horizon to the nearest position of the sun below the horizon at which the star is visible before sunrise / after sunset.

f = observer's geographical latitude, and by extension = Declination of the zenith point.

As noted above, converting f into ecliptic form (B) not only gives the latitude of the zenith point from the ecliptic, but more importantly the angle below the ecliptic, between it and the horizon (= 90˚ - B, since zenith-to-horizon = 90˚).  
This in turn is equal to the angle opposite (angle B), in the triangle bounded by the horizon above (side a) and side $L_1$ (side c) below, with side b (side b) opposite enclosing a right-angle (angle C) with the horizon. By use of the sine-formula:

$$\frac{\sin B}{\sin b} = \frac{\sin C}{\sin c}$$

$$\frac{\sin(90˚ - B)}{\sin b} = \frac{\sin 90˚}{\sin L_1}$$

$$\frac{\sin L_1}{\sin b} = \frac{\sin 90˚}{\sin(90˚ - B)}$$

$$\sin L_1 = \frac{\sin 90˚ \cdot \sin b}{\sin(90˚ - B)}$$

Since $\sin 90˚ = 1$, and $\sin(90˚ - B) = \cos B$:

$$\sin L_1 = \frac{\sin b}{\cos B}$$

hence, $L_1 = \sin^{-1}(\sin b / \cos B)$.

24i)  $L = 10°.3$
$L_1 = 13°.2$
$S = L + L_1 + 90° = 113°.5$

We are now in **ecliptic** coordinates, with the sun on the meridian and of course on the ecliptic, while Sirius is rising or setting.

The angular distance **along the ecliptic** from the sun on the meridian to its position on the horizon is 90°. (This is not its zenith distance, but a measure of how the horizon intersects with the great circle of the ecliptic.)

The ecliptic longitude is measured eastwards from a. So moving from the meridian to the point of rising increases the longitude of the sun, while moving from the meridian to the point of setting decreases it. Therefore, for HR and AR we must add 90° to move the sun to the rising point, but subtract 90° to move it to the setting point. The same applies to $L_1$.

$L + L_1 + 90°$ = the ecliptic longitude of the sun at the time of the heliacal rising of the star. From this the date of the phenomenon can be ascertained.

NB: these are ecliptic coordinates, which have to be converted to equatorial.

REFERENCES

Aujac, G. ed. 1975. *Géminos, Introduction aux Phénomènes*, Paris: Les Belles Lettres.

Bickerman, E. J. 1980. *Chronology of the Ancient World*. London: Thames and Hudson. Revised ed.

Boll, F. 1909. 'Fixsterne,' in A. Pauly, G. Wissowa and W. Kroll, *Real-Encyclopedie des klassischen Altertumwissenschaft*. Stuttgart: 6. 2429-30.

Bömer, F. ed. 1957–58. *P. Ovidius Naso: Die Fasten*. 2 vols. Heidelberg: Winter.

Bowen, A. C. and B. R. Goldstein. 1988. 'Meton of Athens and Astronomy in the Late Fifth Century B.C.,' in E. Leichty, M. de J. ElliS and P. Gerardi, eds., *A Scientific Humanist: Studies in Memory of Abraham Sachs* (Philadelphia,) 39–81.

Buxton, B. and R. Hannah. 2005. 'OGIS 458, the Augustan Calendar, and the Succession', in C. Deroux, ed., *Studies in Latin Literature and Roman History*, volume XII. Brussels: Latomus, 290-306.

Dicks, D. R. 1970. *Early Greek Astronomy to Aristotle*. New York: .

Diels, H. and A. Rehm, 1904. 'Parapegmafragmente aus Milet', *Sitzungsberichte der königlich Preussischen Akademie der Wissenschaften, Philosophisch-historischen*

*Klasse* 23: 92-111.

Evans, J. 1998. *The History and Practice of Ancient Astronomy*. New York and Oxford: Oxford University Press.

Evans, J. and J. L. Berggren. 2006. Geminos's *Introduction to the Phenomena*. Princeton and Oxford: Princeton University Press.

Fantham, R. E. 1998. *Ovid: Fasti IV*. Cambridge: Cambridge University Press.

Fox, M. 2004. 'Stars in the *Fasti*: Ideler (1825) and Ovid's astronomy revisited,' *American Journal of Philology* 125: 91–133.

Frazer, J. G. ed. and trans. 1929. *Publii Ovidii Nasonis: Fastorum libri sex*. 5 vols. London: Macmillan.

Gee, E. 2000. Ovid, *Aratus and Augustus: Astronomy in Ovid's Fasti* (Cambridge: Cambridge University Press.

Ginzel, F. K. 1911. *Handbuch der mathematischen und technischen Chronologie*. vol. 2. Leipzig: J.C. Hinrichs'sche Buchhandlung..

Hannah, R. 1997a. 'The Temple of Mars Ultor and 12 May,' *Mitteilungen des Deutschen Archäologischen Instituts, Römische Abteilung* 104: 374–86.

Hannah, R. 1997b. 'Is it a bird? Is it a star? Ovid's kite and the first swallow of spring,' *Latomus* 56: 327–42.

Hannah, R. 2001. 'The moon, the sun and the stars: counting the days and the years', in S. McCready, ed., *The Discovery of Time*. London: MQ Publications, 56–99.

Hannah, R. 2002. 'Euctemon's Parapegma', in C. J. Tuplin and T. E. Rihll (eds), *Science and Mathematics in Ancient Greek Culture*, Oxford: Oxford University Press, 112–32.

Hannah, R. 2005. *Greek and Roman Calendars: Constructions of Time in the Classical World.* London: Duckworth.

Hannah, R. 2009. *Time in Antiquity*. London: Routledge.

Hannah, R. 2013. 'Greek Government and the Organization of Time', in H. Beck ed. *Companion to Ancient Greek Government*. Oxford: Blackwell, 349–65.

Herbert-Brown, G. 1994. *Ovid and the Fasti: An Historical Study*. New York: Oxford University Press.

Ideler, L. 1825. 'Über den astronomischen Theil der Fasti des Ovid,' *Abhandlungen der königlichen Akademie der Wissenschaft zu Berlin aus den Jahren* 1822–23.

Kidd, D. ed. 1997. *Aratus: Phaenomena*. Cambridge, Cambridge University

Press.

Le Bonniec, H. and A. Le Boeuffle. eds. 1972. *Pline l'Ancien, Histoire Naturelle, Livre XVIII*. Paris: Les Belles Lettres.

Lehoux, D. R. 2005. 'The Parapegma Fragments from Miletus,' *Zeitschrift für Papyrologie und Epigraphik* 152: 125–40.

Lehoux, D. R. 2007. *Astronomy, Weather, and Calendars in the Ancient World : Parapegmata and related texts in Classical and Near Eastern Societies* (Cambridge: Cambridge University Press.

McCluskey, S. 1998. *Astronomies and Cultures in Early Medieval Europe*. Cambridge: .

Nagle, B. R. 1995. *Ovid's Fasti: Roman Holidays*. Bloomington: Indiana University Press.

Neugebauer, O. 1975. *A History of Ancient Mathematical Astronomy*. Berlin and New York: Springer-Verlag.

Neugebauer, P. V. 1925. *Tafeln zur astronomischen Chronologie*. vol. III. Leipzig: J.C. Hinrichs.

Newlands, C. 1995. *Playing with Time: Ovid and the Fasti*. Ithaca: Cornell University Press.

Pritchett, W. K. and B. L. van der Waerden. 1961. 'Thucydidean Time-Reckoning and Euctemon's Seasonal Calendar', *Bulletin de Correspondance Héllenique* 85: 17-52.

Rehm, A. 1913. *Griechische Kalender, iii. Das Parapegma des Euktemon*. Heidelberg: Carl Winter.

Rehm, A. 1941. *Parapegmastudien*. Munich: Verlag der Bayerischen Akademie der Wissenschaften.

Rehm, A. 1949. 'Parapegma', in A. Pauly, G. Wissowa and W. Kroll, *Real-Encyclopedie des klassischen Altertumwissenschaft* Stuttgart: Metzler. 18. 1295-1366.

Rochberg, F. 2004. *The Heavenly Writing: Divination, Horoscopy, and Astronomy in Mesopotamian Culture*. Cambridge: .

Salt, A. and E. Boutsikas. 2005. 'Knowing When to Consult the Oracle at Delphi', *Antiquity* 79: 564– 72.

Schaefer, B. 1985. 'Predicting heliacal risings and settings', *Sky and Telescope* 70: 261-3.

van der Waerden, B. L. 1960. 'Greek Astronomical Calendars and their relation to the Athenian Civil Calendar,' *Journal of Hellenic Studies* 80: 168-80.

van der Waerden, B. L. 1984a. 'Greek Astronomical Calendars. I. The Parapegma of Euctemon', *Archive for History of Exact Sciences* 29: 101-14.

van der Waerden, B. L. 1984b. 'Greek Astronomical Calendars. II. Callipus and his Calendar', *Archive for History of Exact Sciences* 29: 115-24.

van der Waerden, B. L. 1984c. 'Greek Astronomical Calendars. III. The Calendar of Dionysius, *Archive for History of Exact Sciences* 29: 125-30.

van der Waerden, B. L. 1985. 'Greek Astronomical Calendars. IV. The Parapegma of the Egyptians and their 'Perpetual Tables'', *Archive for History of Exact Sciences* 32: 95-104.

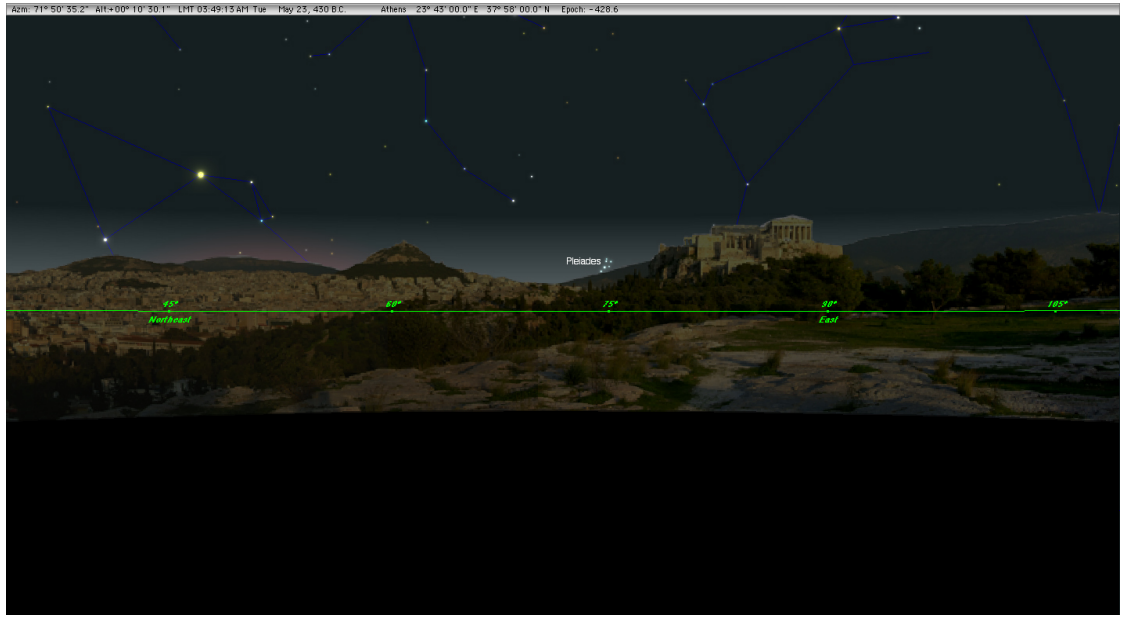

Figure 1

View from the Pnyx, Athens, of the first visibility of the rising Pleiades between the Acropolis and Mt Lykabettos, 16 June 430 BC.

NOTES

[1] Hannah 2009, 2005, 2001.

[2] Hannah 2002: figs. 6.1, 6.2.

[3] Dicks 1970: 84–85.

[4] McCluskey 1998.

[5] Lehoux 2007.

[6] The observations of the *parapegmata* are explicitly called 'rough' by Bowen and Goldstein 1988: 56. More importantly, their accuracy is questioned in one of the standard modern editions of the work in which the *parapegma* of Euktemon is embedded (Aujac 1975), where one will find notes drawing the reader's attention to differences between the dates of observations given in Geminos's compilation and those calculated in modern times. Aujac (1975: 158) expressly includes among the possible causes for these discrepancies 'manque de précision dans l'observation'.

[7] Ideler 1825: 137–69. J. G. Frazer's influential commentary (Frazer 1929) set Ideler's view in concrete for generations; his views are repeated and embellished in recent treatments of the poem, e.g. Bömer 1957–58, Herbert-Brown 1994, Nagle1995, Newlands 1995, Fantham 1998, Gee 2000.

[8] Hannah 1997a, 1997b, Fox 2004.

[9] Diels and Rehm 1904, Rehm 1913, Rehm 1949.

[10] der Waerden 1960, Pritchett and van der Waerden 1961, van der Waerden 1984a, 1984b, 1984c, 1985.

[11] Neugebauer 1975: 595 n.17, Lehoux 2007: 89, 2005: 125–26.

[12] Lehoux 2007; cf. a similar approach for Babylonian astronomical texts in Rochberg 2004.

[13] Hannah 2005: 29–41, 55–58.

[14] Buxton and Hannah 2005.

[15] Hannah 2005: 42–70.

[16] Fox 2004.

[17] Le Bonniec and Le Boeuffle 1972: 264 n. 3.

[18] The classic work based on the assumption of variable angular depression for the sun is Neugebauer 1925. The same assumption informs the dates of star-rise and star-set in the popular tables published by Boll 1909: 2429-30, Ginzel 1911: 517–22, and more recently by Bickerman 1980: 112–14 (a flawed adaptation of Ginzel).

[19] Evans 1998: 249.

[20] Hannah 2009: 116–18.

[21] Recent versions of the Geminos *parapegma*, from which Euktemon's may (in part?) be derived, are those of Aujac 1975, Evans and Berggren 2006, Lehoux 2007.

[22] Voyager 4.5, Carina Software, 865 Ackerman Drive, Danville, CA 94526, USA.

[23] Evans 1998: 197. See also Fox 2004: 104.

[24] Schaefer 1985: 261–3 assumes more stringent parameters, such as the effect of atmospheric extinction; Salt and Boutsikas 2005.

[25] See Hannah 2002: 120–21.

[26] On the meaning of ἐπισημαίνει, which I have translated as 'there is sign of weather', see Evans and Berggren 2006: 230 n.1. They translate it as 'it signifies'; Lehoux 2007: e.g. 233 proposes 'there is a change in the weather'. All understand the usage to refer to a meteorological event.

[27] Salt and Boutsikas 2005.

[28] Aujac omits the manuscript reading, Ἐν δὲ τῇ κε ἡ Εὐκτήμονι Ἀετός ἑσπέριος ἐπιτέλλει [Day 25 (May 19), according to Euktemon Eagle sets in the evening],

because Ἀετός (Eagle) makes no sense astronomically. Αἴξ (Goat) would, but for the sake of consistency, as I have chosen to follow Aujac's text, I have omitted the line.

[29] Kidd 1997: 303.

[30] Hannah 2013: 351–53, with Fig. 23.2.